\documentclass{article}
\def\sleq{\raisebox{-.6ex}{${\textstyle\stackrel{<}{\sim}}$}}

\def\sss{\scriptscriptstyle}
\def\nn{\hspace{2mm}}
\def\sVEV#1{\left\langle #1\right\rangle}
\newcommand{\MeV}{\mbox{\rm MeV}}
\newcommand{\GeV}{\mbox{\rm GeV}}
\newcommand{\eV}{\mbox{\rm eV}}

\title{ \bf Neutrinos in the Family Replicated Gauge Group Model}
\author{Colin D Froggatt, \\[0.2cm]
\it Department of Physics and Astronomy, Glasgow University,\\
\it Glasgow G12 8QQ, Scotland, UK}
\begin{document}
\begin{flushright}
GUTPA/02/10/02\\
\end{flushright}
\vskip .1in
\begin{center}

{\LARGE \bf Neutrinos in the Family Replicated \\
\vspace{6pt}
Gauge Group Model}

\vspace{20pt}

{\bf \large Colin D. Froggatt}

\vspace{6pt}

{ \em Department of Physics and Astronomy\\
 Glasgow University, Glasgow G12 8QQ,
Scotland\\}
\end{center}
\vspace{6pt}
\section*{ }
\begin{center}
{\large\bf Abstract}
\end{center}

We present a discussion of the neutrino mass problem in the 
anti-grand unification theory based on three family replicated 
copies of the Standard Model gauge group (SMG). We consider 
two versions of the theory, with and without right-handed 
neutrinos, and present order of magnitude fits to the 
17 measured quark-lepton mass and mixing angle variables. 
In the model without right-handed neutrinos, based on the 
gauge group $SMG^3\times U(1)_f$ and an isotriplet Higgs,
we obtain a good 4 parameter fit, but with a vacuum oscillation 
solution to the solar neutrino problem. In the second model, 
based on the gauge group $(SMG\times U(1)_{B-L})^3$, we obtain 
a good 5 parameter fit using the usual right-handed neutrino 
see-saw mechanism to generate an LMA-MSW solution to the solar 
neutrino problem. The CHOOZ mixing angle bound is easily 
satisfied in the first model and is close to the predicted 
value in the second model.

\vspace{170pt}

To be published in the 
Proceedings of the International Workshop on 
{\it What comes beyond the Standard Model}, Bled, Slovenia, 
July 2002.

\thispagestyle{empty}
\newpage

\date{}

\section{Introduction}
In this paper, I will update my report \cite{bled2001}
on fermion masses in the Anti-Grand Unification Theory
(AGUT) at the Bled 2001 workshop. There are two versions
of the AGUT model based on three family replicated
copies of the Standard Model (SM) gauge group,
selected according to whether or not each family is
supplemented by a right-handed neutrino. We note that
neither model introduces supersymmetry.
In the absence of right-handed neutrinos, the AGUT gauge
group is $G_1 = SMG^3\times U(1)_f$, where
$SMG \equiv SU(3)\times SU(2) \times U(1)$. With the
inclusion of three right-handed neutrinos, the AGUT
gauge group is extended to
$G_2 = (SMG \times U(1)_{B-L})^3$, where the three
copies of the SM gauge group are supplemented by an
abelian $(B-L)$ (= baryon number minus lepton
number) gauge group for each family.
In each case, the AGUT gauge group $G_1$ ($G_2$)
is the largest anomaly free group \cite{trento}
transforming the
known 45 Weyl fermions (and the additional three
right-handed neutrinos for $G_2$) into each other
unitarily, which does NOT unify the irreducible
representations under the SM gauge group.

Here we present good order of magnitude fits, with four and five
adjustable parameters respectively, to the quark and lepton masses
and mixing angles for the two versions of the AGUT model. In each
case the fit to the charged fermion masses and quark mixings is
arranged to essentially reproduce the original three parameter
AGUT fit \cite{bled2001}. It is necessary to introduce a new mass
scale into the theory, in order to obtain realistic neutrino
masses. For the $G_1 = SMG^3\times U(1)_f$ model we introduce a
weak isotriplet Higgs field  and obtain a vacuum oscillation
solution to the solar neutrino problem, whereas for the $G_2 =
(SMG \times U(1)_{B-L})^3$ model we introduce the usual see-saw
mass scale for the right-handed neutrinos and obtain a large
mixing angle (LMA) MSW solution. During the last year, further
data \cite{Sudbury} from the Sudbury Neutrino Observatory have
confirmed that the LMA-MSW solution is strongly favoured, with the
Vacuum Oscillation and LOW solutions now allowed at the $3 \sigma$
level while the SMA-MSW solution seems to be completely ruled out.
We refer to \cite{Nir} for a recent review of the phenomenology of
neutrino physics.

\section{The $SMG^3\times U(1)_f$ Model}

The usual SM gauge group is identified as the diagonal subgroup of
$SMG^3$ and the AGUT gauge group $SMG^3\times U(1)_f$ is broken
down to this subgroup by four Higgs fields $S$, $W$, $T$ and
$\xi$. Thus, for example, the SM weak hypercharge $y/2$ is given
by the sum of the weak hypercharge quantum numbers $y_i/2$ for the
three proto-families:
\begin{equation}
\frac{y}{2} = \frac{y_1}{2} + \frac{y_2}{2} + \frac{y_3}{2}
\end{equation}
The spontaneously broken chiral AGUT
gauge quantum numbers of the quarks and leptons protect
the charged fermion masses and generate a mass hierarchy
for them \cite{fn} in terms of the Higgs field vacuum
expectation values (VEVs). However the VEV of the
Higgs field $S$ is taken to be unity in
fundamental (Planck) mass units. Thus only the VEVs of
the other three Higgs fields are used as free parameters,
in the order of magnitude fit to the effective SM
Yukawa coupling matrices $Y_U$, $Y_D$ and $Y_E$ for
the quarks and charged leptons \cite{bled2001}.

In this model, the large $\nu_{\mu}-\nu_{\tau}$
mixing required for atmospheric neutrino oscillations
is generated by introducing a large off-diagonal
element, $(Y_E)_{23} \sim (Y_E)_{33}$, in the charged
lepton Yukawa coupling matrix. This is achieved by
introducing a new Higgs field $\psi$ with a VEV equal
to unity in Planck units and the following set of
$U(1)$ gauge charges:
\begin{equation}
\vec{Q}_{\psi}  =  3\vec{Q}_{\xi} + 2\vec{Q}_W + 4\vec{Q}_T
 =  \left ( \frac{1}{2}, -\frac{13}{6}, \frac{5}{3},
-\frac{16}{3} \right),
\end{equation}
Here we express the abelian gauge charges in the model as a
charge vector $\vec{Q} = (y_1/2, y_2/2, y_3/2, Q_f)$.

We then have the Yukawa matrices for the quarks:
\begin{equation}
Y_U  =  \left(\begin{array}{ccc} WT^2\xi^2 & WT^2\xi & W^2T\xi \\
                                   WT^2\xi^3 & WT^2    & W^2T    \\
                                   \xi^3     & 1       & WT \end{array}
          \right), \quad
Y_D  =  \left(\begin{array}{ccc} WT^2\xi^2 & WT^2\xi & T^3\xi  \\
                                   WT^2\xi   & WT^2    & T^3     \\
                                   W^3T & W^3T\xi  & WT \end{array}
          \right)
\end{equation}
and the charged lepton Yukawa matrix:
\begin{equation}
Y_E  =  \left(\begin{array}{ccc} WT^2\xi^2 & W^3T^2 & W^2T\xi^3 \\
                                   WT^2\xi^5 & WT^2    & W\xi \\
                                   W^4\xi^2 & W^3T\xi  & WT \end{array}
          \right).
\label{eq:yukabel2}
\end{equation}
We still obtain a good order of magnitude phenomenology for the charged
fermion masses and quark mixing angles, similar to the original
AGUT fit \cite{bled2001}, with the following VEVs in Planck units:
\begin{equation}
<W> = 0.179, \qquad <T> = 0.071 \qquad <\xi> = 0.099.
\end{equation}

The unitary matrix $U_E$ which diagonalises $Y_EY_E^{\dagger}$
is then given by
\begin{equation}
\left( \begin{array}{ccc}
1 & \frac{W\xi^4}{T^3} & W \xi^3  \\
-\frac{W\xi^4}{T^3\sqrt{1+\frac{\xi^2}{T^2}}} &
     \frac{1}{\sqrt{1+\frac{\xi^2}{T^2}}}
     & \frac{\xi}{T\sqrt{1+\frac{\xi^2}{T^2}}} \\
\frac{W\xi^5}{T^4\sqrt{1+\frac{\xi^2}{T^2}}}
     & -\frac{\xi}{T\sqrt{1+\frac{\xi^2}{T^2}}} &
    \frac{1}{\sqrt{1+\frac{\xi^2}{T^2}}}
    \end{array}\right)\\
 \sim  \left(\hspace{-2pt} \begin{array}{ccc}
1 & 0.05& 1.7\times 10^{-4} \\
- 0.03& 0.58 & 0.81 \\
0.04 & - 0.81 & 0.58 \end{array} \hspace{-2pt}\right)
\end{equation}
As we can see from the structure of $U_E$, we naturally
obtain the large $\mu-\tau$ mixing required for the atmospheric
neutrinos. We can now obtain suitable mixing and a
suitable hierarchy of neutrino masses for a vacuum oscillation
solution to the solar neutrino problem, by making the following
choice of charges for a weak iso-triplet
Higgs field, $\Delta$.
\begin{equation}
\vec{Q}_{\Delta} = (\frac{1}{2}, \frac{1}{3}, -\frac{5}{6}, 0).
\end{equation}
We then have the neutrino mass matrix,
\begin{equation}
M_{\nu} \sim <\Delta^0> \left (
\begin{array}{ccc}
W\xi^6 & W\xi^3 & WT\xi^2\\
W \xi^3 & W & WT\xi\\
WT\xi^2 & WT\xi & WT^2\xi \end{array} \right)
\end{equation}
This has the hierarchy,
\begin{eqnarray}
\Delta m^2_{12} & \sim & \Delta m^2_{23},\\
\frac{\Delta m^2_{13}}{\Delta m^2_{12}} & \sim & 2 T^3 \xi^3 \sim
7 \times 10^{-7},
\end{eqnarray}
which is just suitable for the atmospheric neutrinos and the
vacuum oscillation solution to the solar neutrino problem. The
electron neutrino mixing is also large enough for the vacuum
oscillation solution to the solar neutrino problem, as we
can see from the matrix $U_{\nu}$ which diagonalises $M_{\nu}$:
\begin{equation}
U_{\nu} \sim \left (
\begin{array}{ccc} \frac{1}{\sqrt{2}}(1 + \frac{T}{4\xi}) & \xi^3
&
\frac{1}{\sqrt{2}}(1 - \frac{T}{4\xi}) \\
\frac{T\xi}{\sqrt{2}}(1 - \frac{T}{4\xi}) & 1 &
-\frac{T\xi}{\sqrt{2}}(1 + \frac{T}{4\xi}) \\
-\frac{1}{\sqrt{2}}(1 - \frac{T}{4\xi}) & T\xi &
\frac{1}{\sqrt{2}}(1 + \frac{T}{4\xi})
\end{array} \right).
\end{equation}
Hence we have the lepton mixing matrix,
\begin{eqnarray}
U = U_E^{\dagger} U_{\nu} & \sim & \left( \begin{array}{ccc}
\frac{1}{\sqrt{2}}(1 + \frac{T}{4\xi}) &
\frac{W \xi^4}{T^3 \sqrt{1+\frac{\xi^2}{T^2}}} &
\frac{1}{\sqrt{2}(1 - \frac{T}{4\xi})}\\
\frac{\xi(1-\frac{T}{4\xi})}{\sqrt{2}T\sqrt{1+\frac{\xi^2}{T^2}}}
& \frac{1}{\sqrt{1+\frac{\xi^2}{T^2}}}
    & -\frac{\xi(1+\frac{T}{4\xi})}{T\sqrt{2(1+\frac{\xi^2}{T^2})}}\\
-\frac{1-\frac{T}{4\xi}}{\sqrt{2(1+\frac{\xi^2}{T^2})}} &
\frac{\xi}{T \sqrt{1+\frac{\xi^2}{T^2}}} &
\frac{1+\frac{T}{4\xi}}{\sqrt{2(1+\frac{\xi^2}{T^2})}}
\end{array} \right)\\
& \sim & \left( \begin{array}{ccc}
0.83 & 2.8 \times 10^{-2} & 0.58\\
0.47 & 0.58 & -0.68\\
-0.34 & 0.813 & 0.49
\end{array} \right).
\end{eqnarray}
If we take $<\Delta^0> \sim 0.18 \ \mbox{eV}$, then we have
\begin{eqnarray}
m_1 & \sim & <\Delta^0> \left(-W T \xi^2 + \frac{T^2 \xi W}{2}
\right)
\sim - 1.4 \times 10^{-5} \ \mbox{eV} \nonumber \\
m_2 & \sim & <\Delta^0> W \sim 3.2 \times 10^{-2}
\ \mbox{eV} \nonumber \\
m_3 & \sim & <\Delta^0> \left(W T \xi^2 + \frac{T^2 \xi W}{2}
\right) \sim  3 \times 10^{-5} \ \mbox{eV}.
\end{eqnarray}
The $\Delta m^2$ and mixing angle for the solar neutrinos
are given by
\begin{equation}
\Delta m^2_{13} \sim 7 \times 10^{-10} \ \mbox{eV}^2,
\qquad \sin^2 2\theta_{\odot} \sim 4 U_{e1}^2 U_{e3}^2 \sim 0.93
\end{equation}
which are compatible with vacuum oscillations for the solar
neutrinos. Similarly the $\Delta m^2$ and mixing angle for
atmospheric neutrinos are given by,
\begin{equation}
\Delta m^2_{23} \sim 1 \times 10^{-3} \ \mbox{eV}^2,
\qquad \sin^2 2\theta_{atm} \sim 0.93.
\end{equation}
Hence we can see that we have large (but not maximal) mixing for
both the solar and atmospheric neutrinos. We also note that the 
CHOOZ electron survival probability bound is readily satisfied by
$U_{e2} \sim 0.028 < 0.16$, which is the relevant mixing matrix element
since $\Delta m_{12}^2 \sim \Delta m_{23}^2 \gg \Delta m_{13}^2$.
Thus we obtain a good order of magnitude fit (agreeing with the 
data to within a factor of 2) to the 17 measured fermion mass and 
mixing angle variables with just 4 free parameters ($W$, $T$, 
$\xi$ and $\Delta^0$), but assuming a vacuum oscillation solution to 
the solar neutrino problem.

We note that we have really only used the abelian gauge quantum
numbers to generate a realistic spectrum of fermion masses; the
non-abelian representations are determined by imposing the usual
SM charge quantisation rule for each of the SMG factors in the
gauge group. Furthermore two of the $U(1)$s (more precisely
two linear combinations of the $U(1)$s) in the gauge group are
spontaneously broken by the Higgs fields $S$ and $\psi$, which
have VEVs $<S> = <\psi> = 1$. Hence these $U(1)$s play
essentially no part in obtaining the spectrum of fermion
masses and mixings. This means that we can construct a model
based on the gauge group $SMG \times U(1)^{\prime}$ with the same
fermion spectrum as above. However it turns out \cite{mg}
that some of the quarks and leptons must have extremely
large (integer) $U(1)^{\prime}$ charges, making
this reduced model rather unattractive. Also three Higgs fields
$W$, $T$ and $\xi$ are responsible for the spontaneous breakdown
to the SM gauge group, $SMG \times U(1)^{\prime} \rightarrow
SMG$, and they have large relatively prime $U(1)^{\prime}$
charges. So we prefer the better motivated $SMG^3\times U(1)_f$
AGUT model.

\section{The $(SMG \times U(1)_{B-L})^3$ Model}

In this extended AGUT model we introduce a right-handed neutrino
and a gauged $B-L$ charge for each family with the associated
abelian gauge groups $U(1)_{\sss B-L,i}$ ($i=1,2,3$). The $U(1)_f$
abelian factor of the $SMG \times U(1)_f$ model in the previous
section gets absorbed as a linear combination of the $B-L$ charge
and the weak hypercharge abelian gauge groups for the different
families (or generations). It is these 6 abelian gauge charges
which are responsible for generating the fermion mass hierarchy
and we list their values in Table \ref{Table1} for the 48 Weyl
proto-fermions in the model.
\begin{table}[!ht]
\caption{All $U(1)$ quantum charges for the proto-fermions in the
$(SMG \times U(1)_{B-L})^3$ model.} \vspace{3mm} \label{Table1}
\begin{center}
\begin{tabular}{|c||c|c|c|c|c|c|} \hline
& $SMG_1$& $SMG_2$ & $SMG_3$ & $U_{\sss B-L,1}$ & $U_{\sss B-L,2}$
& $U_{\sss B-L,3}$ \\ \hline\hline
$u_L,d_L$ &  $\frac{1}{6}$ & $0$ & $0$ & $\frac{1}{3}$ & $0$ & $0$ \\
$u_R$ &  $\frac{2}{3}$ & $0$ & $0$ & $\frac{1}{3}$ & $0$ & $0$ \\
$d_R$ & $-\frac{1}{3}$ & $0$ & $0$ & $\frac{1}{3}$ & $0$ & $0$ \\
$e_L, \nu_{e_{\sss L}}$ & $-\frac{1}{2}$ & $0$ & $0$ & $-1$ & $0$ & $0$ \\
$e_R$ & $-1$ & $0$ & $0$ & $-1$ & $0$ & $0$ \\
$\nu_{e_{\sss R}}$ &  $0$ & $0$ & $0$ & $-1$ & $0$ & $0$ \\ \hline
$c_L,s_L$ & $0$ & $\frac{1}{6}$ & $0$ & $0$ & $\frac{1}{3}$ & $0$ \\
$c_R$ &  $0$ & $\frac{2}{3}$ & $0$ & $0$ & $\frac{1}{3}$ & $0$ \\
$s_R$ & $0$ & $-\frac{1}{3}$ & $0$ & $0$ & $\frac{1}{3}$ & $0$\\
$\mu_L, \nu_{\mu_{\sss L}}$ & $0$ & $-\frac{1}{2}$ & $0$ & $0$ & $-1$ &
$0$\\ $\mu_R$ & $0$ & $-1$ & $0$ & $0$  & $-1$ & $0$ \\
$\nu_{\mu_{\sss R}}$ &  $0$ & $0$ & $0$ & $0$ & $-1$ & $0$ \\ \hline
$t_L,b_L$ & $0$ & $0$ & $\frac{1}{6}$ & $0$ & $0$ & $\frac{1}{3}$ \\
$t_R$ &  $0$ & $0$ & $\frac{2}{3}$ & $0$ & $0$ & $\frac{1}{3}$ \\
$b_R$ & $0$ & $0$ & $-\frac{1}{3}$ & $0$ & $0$ & $\frac{1}{3}$\\
$\tau_L, \nu_{\tau_{\sss L}}$ & $0$ & $0$ & $-\frac{1}{2}$ & $0$ & $0$ &
$-1$\\ $\tau_R$ & $0$ & $0$ & $-1$ & $0$ & $0$ & $-1$\\
$\nu_{\tau_{\sss R}}$ &  $0$ & $0$ & $0$ & $0$ & $0$ & $-1$ \\
\hline \hline
\end{tabular}
\end{center}
\end{table}
The see-saw scale for the right-handed neutrinos is introduced via
the VEV of a new Higgs field $\phi_{\sss SS}$. However, in order
to get an LMA-MSW solution to the solar neutrino problem, we have
to replace \cite{fnt} the AGUT Higgs fields $S$ and $\xi$ by two
new Higgs fields $\rho$ and $\omega$. The abelian gauge quantum
numbers of the new system of Higgs fields for the $(SMG \times
U(1)_{B-L})^3$ model are given in Table \ref{qc}.
\begin{table}[!th]
\caption{All $U(1)$ quantum charges of the Higgs fields in the
$(SMG \times U(1)_{B-L})^3$ model.} \vspace{3mm} \label{qc}
\begin{center}
\begin{tabular}{|c||c|c|c|c|c|c|} \hline
& $SMG_1$& $SMG_2$ & $SMG_3$ & $U_{\sss B-L,1}$ & $U_{\sss B-L,2}$
& $U_{\sss B-L,3}$ \\ \hline\hline
$\omega$ & $\frac{1}{6}$ & $-\frac{1}{6}$ & $0$ & $0$ & $0$ & $0$\\
$\rho$ & $0$ & $0$ & $0$ & $-\frac{1}{3}$ & $\frac{1}{3}$ & $0$\\
$W$ & $0$ & $-\frac{1}{2}$ & $\frac{1}{2}$ & $0$ & $-\frac{1}{3}$
& $\frac{1}{3}$ \\
$T$ & $0$ & $-\frac{1}{6}$ & $\frac{1}{6}$ & $0$ & $0$ & $0$\\
$\phi_{\sss WS}$ & $0$ & $\frac{2}{3}$ & $-\frac{1}{6}$ & $0$
& $\frac{1}{3}$ & $-\frac{1}{3}$ \\
$\phi_{\sss SS}$ & $0$ & $1$ & $-1$ & $0$ & $2$ & $0$ \\
\hline
\end{tabular}
\end{center}
\end{table}

As can be seen from Table~\ref{qc}, the fields $\omega$ and $\rho$
have only non-trivial quantum numbers with respect to the first
and second families. This choice of quantum numbers makes it
possible to express a fermion mass matrix element involving the
first family in terms of the corresponding element involving the
second family, by the inclusion of an appropriate product of
powers of $\rho$ and $\omega$. With the system of quantum numbers
in Table~\ref{qc} one can easily evaluate, for a given mass matrix
element, the numbers of Higgs field VEVs of the different types
needed to perform the transition between the corresponding left-
and right-handed Weyl fields. The results of calculating the
products of Higgs fields needed, and thereby the order of
magnitudes of the mass matrix elements in our model, are presented
in the following mass matrices (where, for clarity, we distinguish
between Higgs fields and their hermitian conjugates):

\noindent the up-type quarks:
\begin{eqnarray}
M_{\sss U} \simeq \frac{\sVEV{(\phi_{\sss\rm WS})^\dagger}}
{\sqrt{2}}\hspace{-0.1cm}
\left(\!\begin{array}{ccc}
        (\omega^\dagger)^3 W^\dagger T^2
        & \omega \rho^\dagger W^\dagger T^2
        & \omega \rho^\dagger (W^\dagger)^2 T\\
        (\omega^\dagger)^4 \rho W^\dagger T^2
        &  W^\dagger T^2
        & (W^\dagger)^2 T\\
        (\omega^\dagger)^4 \rho
        & 1
        & W^\dagger T^\dagger
\end{array} \!\right)\label{M_U}
\end{eqnarray}
\noindent
the down-type quarks:
\begin{eqnarray}
M_{\sss D} \simeq \frac{\sVEV{\phi_{\sss\rm WS}}}{\sqrt{2}}
\hspace{-0.1cm}
\left (\!\begin{array}{ccc}
        \omega^3 W (T^\dagger)^2
      & \omega \rho^\dagger W (T^\dagger)^2
      & \omega \rho^\dagger T^3 \\
        \omega^2 \rho W (T^\dagger)^2
      & W (T^\dagger)^2
      & T^3 \\
        \omega^2 \rho W^2 (T^\dagger)^4
      & W^2 (T^\dagger)^4
      & W T
                        \end{array} \!\right) \label{M_D}
\end{eqnarray}
\noindent %
the charged leptons:
\begin{eqnarray}
M_{\sss E} \simeq \frac{\sVEV{\phi_{\sss\rm WS}}}{\sqrt{2}}
\hspace{-0.1cm}
\left(\hspace{-0.1 cm}\begin{array}{ccc}
    \omega^3 W (T^\dagger)^2
  & (\omega^\dagger)^3 \rho^3 W (T^\dagger)^2
  & (\omega^\dagger)^3 \rho^3 W^4 (T^\dagger)^5  \\
    \omega^6 (\rho^\dagger)^3  W (T^\dagger)^2
  &   W (T^\dagger)^2
  &   W^4 (T^\dagger)^5 \\
    \omega^6 (\rho^\dagger)^3  (W^\dagger)^2 T^4
  & (W^\dagger)^2 T^4
  & WT
\end{array} \hspace{-0.1cm}\right) \label{M_E}
\end{eqnarray}
\noindent
the Dirac neutrinos:
\begin{eqnarray}
M^D_\nu \simeq \frac{\sVEV{(\phi_{\sss\rm WS})^\dagger}}
{\sqrt{2}}\hspace{-0.1cm}
\left(\hspace{-0.1cm}\begin{array}{ccc}
        (\omega^\dagger)^3 W^\dagger T^2
        & (\omega^\dagger)^3 \rho^3 W^\dagger T^2
        & (\omega^\dagger)^3 \rho^3 W^2 (T^\dagger)^7 \\
        (\rho^\dagger)^3 W^\dagger T^2
        &  W^\dagger T^2
        & W^2 (T^\dagger)^7 \\
        (\rho^\dagger)^3 (W^\dagger)^4 T^8
        &   (W^\dagger)^4 T^8
        & W^\dagger T^\dagger
\end{array} \hspace{-0.1 cm}\right)\label{Mdirac}
\end{eqnarray}
\noindent %
and the Majorana (right-handed) neutrinos:
\begin{eqnarray}
M_R \simeq \sVEV{\phi_{\sss\rm SS}}\hspace{-0.1cm}
\left (\hspace{-0.1 cm}\begin{array}{ccc}
(\rho^\dagger)^6 T^6
& (\rho^\dagger)^3 T^6
& (\rho^\dagger)^3 W^3 (T^\dagger)^3 \\
(\rho^\dagger)^3 T^6
& T^6 & W^3 (T^\dagger)^3 \\
(\rho^\dagger)^3 W^3 (T^\dagger)^3 & W^3 (T^\dagger)^3
& W^6 (T^\dagger)^{12}
\end{array} \hspace{-0.1 cm}\right ) \label{Mmajo}
\end{eqnarray}

Then the light neutrino mass matrix -- effective left-left
transition Majorana mass matrix -- can be obtained via the see-saw
mechanism~\cite{seesaw}:
\begin{equation}
  \label{eq:meff}
  M_{\rm eff} \! \approx \! M^D_\nu\,M_R^{-1}\,(M^D_\nu)^T\nn.
\end{equation}
with an appropriate renormalisation group running from the Planck 
scale to the see-saw scale and then to the electroweak scale. The
experimental quark and lepton masses and mixing angles in Table
\ref{convbestfit} can now be fitted, by varying just 5 Higgs field
VEVs and averaging over a set of complex order unity random 
numbers, which multiply all the independent mass matrix
elements. The best fit is obtained with the following values for
the VEVs:
\begin{eqnarray}
\label{eq:VEVS} &&\sVEV{\phi_{\sss SS}}=5.25\times10^{15}~\GeV\nn,
\nn\sVEV{\omega}=0.244\nn, \nn\sVEV{\rho}=0.265\nn,\nonumber\\
&&\nn\sVEV{W}=0.157\nn, \nn\sVEV{T}=0.0766\nn,
\end{eqnarray}
where, except for the Higgs field $\sVEV{\phi_{\sss SS}}$, the
VEVs are expressed in Planck units. The resulting 5 parameter 
order of magnitude fit, with an LMA-MSW solution to the 
solar neutrino problem, is presented
in Table \ref{convbestfit}.

\begin{table}[!t]
\caption{Best fit to conventional experimental data. All masses
are running masses at $1~\GeV$ except the top quark mass which is
the pole mass.}
\begin{displaymath}
\begin{array}{|c|c|c|}
\hline\hline
 & {\rm Fitted} & {\rm Experimental} \\ \hline
m_u & 4.4~\MeV & 4~\MeV \\
m_d & 4.3~\MeV & 9~\MeV \\
m_e & 1.6~\MeV & 0.5~\MeV \\
m_c & 0.64~\GeV & 1.4~\GeV \\
m_s & 295~\MeV & 200~\MeV \\
m_{\mu} & 111~\MeV & 105~\MeV \\
M_t & 202~\GeV & 180~\GeV \\
m_b & 5.7~\GeV & 6.3~\GeV \\
m_{\tau} & 1.46~\GeV & 1.78~\GeV \\
V_{us} & 0.11 & 0.22 \\
V_{cb} & 0.026 & 0.041 \\
V_{ub} & 0.0027 & 0.0035 \\ \hline
\Delta m^2_{\odot} & 9.0 \times 10^{-5}~\eV^2 &  5.0 \times 10^{-5}~\eV^2 \\
\Delta m^2_{\rm atm} & 1.7 \times 10^{-3}~\eV^2 &  2.5 \times 10^{-3}~\eV^2\\
\tan^2\theta_{\odot} &0.26 & 0.34\\
\tan^2\theta_{\rm atm}& 0.65 & 1.0\\
\tan^2\theta_{\rm chooz}  & 2.9 \times 10^{-2} & \sleq~2.6 \times 10^{-2}\\
\hline\hline
\end{array}
\end{displaymath}
\label{convbestfit}
\end{table}

Transforming from $\tan^2\theta$ variables to $\sin^22\theta$
variables, our predictions for the neutrino mixing angles become:
\begin{equation}
  \label{eq:sintan}
 \sin^22\theta_{\odot} = 0.66\nn, \quad
 \sin^22\theta_{\rm atm} = 0.96\nn, \quad
 \sin^22\theta_{\rm chooz} = 0.11\nn.
\end{equation}
Note that our fit to the CHOOZ mixing angle lies close to the 
$2\sigma$ Confidence Level
experimental bound. We also give here our predicted hierarchical
left-handed neutrino masses ($m_i$) and the right-handed neutrino
masses ($M_i$) with mass eigenstate indices ($i=1,2,3$):
\begin{equation}
m_1 =  1.4\times10^{-3}~~\eV\nn, \quad
M_1 =  1.0\times10^{6}~~\GeV\nn,
\label{eq:neutrinomass1}
\end{equation}
\begin{equation}
m_2 =  9.6\times10^{-3}~~\eV\nn, \quad
M_2 =  6.1\times10^{9}~~\GeV\nn,
\label{eq:neutrinomass2}
\end{equation}
\begin{equation}
m_3 =  4.2\times10^{-2}~~\eV\nn, \quad
M_3 =  7.8\times10^{9}~~\GeV\nn.
\label{eq:neutrinomass3}
\end{equation}

\section*{Acknowledgements}

I should again like to thank my collaborators, Mark Gibson,
Holger Bech Nielsen and Yasutaka Takanishi, for many
discussions.

\end{document}